\documentclass{llncs}

\usepackage{a4,a4wide}
\usepackage{url}

\usepackage{latexsym}
\usepackage{xspace}
\usepackage{color}

\usepackage{listings}
\lstnewenvironment{curry}{\lstset{backgroundcolor=\color[rgb]{0.9,0.9,0.9}}}{}
\lstnewenvironment{haskell}{}{}
\newcommand{\listline}{\vrule width0pt depth1.75ex}

\newcommand{\code}[1]{\mbox{\small\texttt{#1}}}
\newcommand{\ccode}[1]{``\code{#1}''}
\newcommand{\us}{\char95\xspace} 

\begin{document}
\sloppy

\title{Implementing Equational Constraints\\ in a Functional Language}

\author{
Bernd Bra{\ss}el
\and
Michael Hanus
\and
Bj{\"o}rn Peem{\"o}ller
\and
Fabian Reck
}
\institute{
Institut f\"ur Informatik, CAU Kiel, D-24098 Kiel, Germany \\
\email{\{bbr|mh|bjp|fre\}@informatik.uni-kiel.de}
}

\maketitle

\begin{abstract}
KiCS2 is a new system
to compile functional logic programs of the source language Curry
into purely functional Haskell programs.
The implementation is based on the idea to represent
the search space as a data structure and logic variables
as operations that generate their values.
This has the advantage that one can apply various,
and in particular, complete search strategies to compute solutions.
However, the generation of all values for logic variables
might be inefficient for applications that
exploit constraints on partially known values.
To overcome this drawback, we propose new techniques
to implement equational constraints in this framework.
In particular, we show how unification modulo function evaluation
and functional patterns can be added without sacrificing
the efficiency of the kernel implementation.
\end{abstract}

\section{Introduction}
\label{sec:Introduction}

Functional logic languages combine the most important
features of functional and logic programming in a single language
(see \cite{AntoyHanus10CACM,Hanus07ICLP} for recent surveys).
In particular, they provide higher-order functions and demand-driven
evaluation from functional programming together with logic programming features
like non-deterministic search and computing with partial information
(logic variables).
This combination
has led to new design patterns \cite{AntoyHanus02FLOPS,AntoyHanus11WFLP}
and better abstractions for application programming, but it also gave rise
to new implementation challenges.

Previous implementations of functional logic languages
can be classified into three categories:
\begin{enumerate}
 \item designing new abstract machines appropriately supporting
these operational features and implementing them in some (typically,
imperative) language, like C \cite{Lux99FLOPS}
or Java \cite{AntoyHanusLiuTolmach05,HanusSadre99JFLP},

 \item compilation into logic languages like Prolog and reusing
the existing backtracking implementation for non-deterministic
search as well as logic variables and unification for computing with partial
information \cite{AntoyHanus00FROCOS,Lopez-FraguasSanchez-Hernandez99}, or

 \item compilation into non-strict functional languages like Haskell
and reusing the implementation
of lazy evaluation and higher-order functions
\cite{BrasselHuch07,BrasselHuch09}.
\end{enumerate}
The latter approach requires the implementation
of non-deterministic computations in a deterministic language
but has the advantage
that the explicit handling of non-determinism allows for
various search strategies like depth-first, breadth-first, parallel,
or iterative deepening instead of committing to a fixed (incomplete)
strategy like backtracking \cite{BrasselHuch07}.

In this paper we consider KiCS2 \cite{BrasselHanusPeemoellerReck11},
a new system that compiles functional logic programs of the
source language Curry \cite{Hanus06Curry}
into purely functional Haskell programs.
We have shown in \cite{BrasselHanusPeemoellerReck11}
that this implementation can compete with or outperform
other existing implementations of Curry.
KiCS2 is based on the idea to represent
the search space, i.e., all non-deterministic results of a computation,
as a data structure that can be traversed by operations
implementing various strategies.
Furthermore, logic variables are replaced by generators,
i.e., operations that non-deterministically evaluate to
all possible ground values of the type of the logic variable.
It has been shown \cite{AntoyHanus06ICLP}
that computing with logic variables by narrowing \cite{Reddy85,Slagle74}
and computing with generators by rewriting
are equivalent, i.e., compute the same values.
Although this implementation technique is correct \cite{Brassel11Thesis},
the generation of all values for logic variables
might be inefficient for applications that
exploit constraints on partial values.
For instance, in Prolog the equality constraint \ccode{X=c(a)}
is solved by instantiating the variable \code{X} to \code{c(a)},
but the equality constraint \ccode{X=Y} is solved by binding \code{X}
to \code{Y} without enumerating any values for \code{X} or \code{Y}.
Therefore, we propose in this paper new techniques
to implement equational constraints in the framework of KiCS2
(note that, in contrast to Prolog, unification is performed
modulo function evaluation).
Furthermore, we also show how functional patterns \cite{AntoyHanus05LOPSTR},
i.e., patterns containing evaluable operations for more powerful pattern 
matching than in logic or functional languages,
can be implemented in this framework.
We show that both extensions lead to efficiency improvements
without sacrificing the efficiency of the kernel implementation.

In the next section, we review the source language Curry
and the features considered in this paper.
Section~\ref{sec:Compilation} sketches the implementation scheme
of KiCS2.
Sections~\ref{sec:Unification} and~\ref{sec:FuncPatterns}
discuss the extensions to implement unification modulo
functional evaluation and functional patterns, respectively.
Benchmarks demonstrating the usefulness of this scheme
are presented in Sect.~\ref{sec:Benchmarks}
before we conclude in Sect.~\ref{sec:Conclusions}.

\section{Curry Programs}
\label{sec:Curry}

The syntax of the functional logic language Curry \cite{Hanus06Curry}
is close to Haskell \cite{PeytonJones03Haskell},
i.e., type variables and names of defined operations usually
start with lowercase letters and the names of type and data constructors
start with an uppercase letter. The application of $f$
to $e$ is denoted by juxtaposition (``$f~e$'').
In addition to Haskell, Curry allows free (logic)
variables in conditions and right-hand sides of defining rules.
Hence, an operation is defined by conditional rewrite rules of the form:
\begin{equation}
\label{rule}
f~t_1 \ldots t_n \code{~|~} c \code{~=~} e \code{~~where~} vs \code{~free}
\end{equation}
where the \emph{condition} $c$ is optional and
$vs$ is the list of variables occurring in $c$ or $e$ but not in the
\emph{left-hand side} $f~t_1 \ldots t_n$.

In contrast to functional programming and similarly to logic programming,
operations can be defined by overlapping rules so that
they might yield more than one result on the same input.
Such operations are also called \emph{non-deterministic}.
For instance, Curry offers a \emph{choice} operation that is predefined by
the following rules:
\begin{curry}
  x ? _ = x
  _ ? y = y
\end{curry}
Thus, we can define a non-deterministic operation \code{aBool} by
\label{ex:aBool}
\begin{curry}
  aBool = True ? False
\end{curry}
so that the expression \ccode{aBool} has two values:
\code{True} and \code{False}.

If non-deterministic operations are used as arguments in other operations,
a semantical ambiguity might occur. Consider the operations
\begin{curry}
  not True  = False
  not False = True

  xor True  x = not x
  xor False x = x

  xorSelf x = xor x x
\end{curry}
and the expression \ccode{xorSelf aBool}.
If we interpret this program as a term rewriting system,
we could have the reduction
\begin{haskell}
  xorSelf aBool  $\to~$  xor aBool aBool     $\to~$  xor True aBool
                 $\to~$  xor True False      $\to~$  not False        $\to~$ True
\end{haskell}
leading to the unintended result \code{True}.
Note that this result cannot be obtained if we use a strict strategy
where arguments are evaluated prior to the function calls.
In order to avoid dependencies on the evaluation strategies
and exclude such unintended results,
Gonz\'alez-Moreno et al.\ \cite{GonzalezEtAl99} proposed
the rewriting logic CRWL as a logical
(execution- and strategy-independent) foundation for declarative
programming with non-strict and non-deterministic operations.  This
logic specifies the \emph{call-time choice} semantics \cite{Hussmann92}
\label{ctc-semantics},
where values of the arguments of an operation are determined before the
operation is evaluated. This can be enforced in a lazy strategy
by sharing actual arguments.
For instance, the expression above can be lazily evaluated
provided that all occurrences of \code{aBool}
are shared so that all of them reduce either to \code{True} or to \code{False}
consistently.

The condition $c$ in rule (\ref{rule}) typically is a
conjunction of \emph{equational constraints}
of the form \code{$e_1$\,=:=\,$e_2$}.
Such a constraint is satisfiable if
both sides $e_1$ and $e_2$ are reducible to unifiable data terms.
For instance, if the symbol ``\code{++}'' denotes the usual list
concatenation operation, we can define an operation \code{last}
that computes the last element \code{e} of a non-empty list \code{xs}
as follows:
\label{ex:last}
\begin{curry}
  last xs | ys++[e] =:= xs  = e   where ys, e free
\end{curry}
Like in Haskell, most rules defining functions are \emph{constructor-based} 
\cite{ODonnell85}, i.e., in (\ref{rule})
$t_1,\ldots,t_n$ consist of variables and/or data constructor symbols
only.  However, Curry also allows \emph{functional patterns}
\cite{AntoyHanus05LOPSTR}, i.e., $t_i$ might additionally contain calls to
defined operations. For instance, we can also define the last element
of a list by:
\begin{curry}
  last' (xs++[e]) = e
\end{curry}
Here, the functional pattern \code{(xs++[e])}
states that \code{(last' t)} is reducible
to \code{e} provided that the argument \code{t} can be matched against
some value of \code{(xs++[e])} where \code{xs} and \code{e} are
free variables.
By instantiating \code{xs} to arbitrary lists, the value
of \code{(xs++[e])} is any list having \code{e} as its last element.
Functional patterns are a powerful feature to express arbitrary 
selections in term structures.
For instance, they support a straightforward
processing of XML data with incompletely specified
or evolving formats \cite{Hanus11ICLP}.

\section{The Compilation Scheme of KiCS2}
\label{sec:Compilation}

To understand the extensions described in the subsequent sections,
we sketch the translation of Curry programs
into Haskell programs as performed by KiCS2.
More details about this translation scheme can be found in
\cite{BrasselFischer08IFL,BrasselHanusPeemoellerReck11}.

As mentioned in the introduction, the KiCS2 implementation
is based on the explicit representation of non-deterministic results
in a data structure.
This is achieved by extending each data type of the source program
by constructors to represent a choice between two values
and a failure, respectively.
For instance, the data type for Boolean values defined in a Curry program by
\begin{curry}
  data Bool = False | True
\end{curry}
is translated into the Haskell data type\footnote{Actually,
our compiler performs some renamings to avoid conflicts with
predefined Haskell entities and introduces type classes
to resolve overloaded symbols like \code{Choice} and \code{Fail}.}
\begin{haskell}
  data Bool = False | True | Choice ID Bool Bool | Fail
\end{haskell}
The first argument of type \code{ID} of each \code{Choice} constructor
is used to implement the call-time choice semantics
discussed in Sect.~\ref{sec:Curry}.
Since the evaluation of \code{xorSelf aBool} duplicates
the argument operation \code{aBool}, we have to ensure
that both duplicates, which later evaluate to a non-deterministic
choice between two values, yield either \code{True} or \code{False}.
This is obtained by assigning a unique identifier (of type \code{ID})
to each \code{Choice}. The difficulty is to get a unique identifier
on demand, i.e., when some operation evaluates to a \code{Choice}.
Since we want to compile into \emph{purely} functional programs
(in order to enable powerful program optimizations),
we cannot use unsafe features with side effects to generate
such identifiers.
Hence, we pass a (conceptually infinite) set of identifiers,
also called \emph{identifier supply},
to each operation so that a \code{Choice} can pick its unique identifier from
this set.
For this purpose, we assume a type \code{IDSupply},
representing an infinite set of identifiers,
with operations
\begin{haskell}
  initSupply  :: IO IDSupply
  thisID      :: IDSupply -> ID
  leftSupply  :: IDSupply -> IDSupply
  rightSupply :: IDSupply -> IDSupply
\end{haskell}
The operation \code{initSupply} creates such a set (at the beginning
of an execution),
the operation \code{thisID} takes some identifier from this set, and
\code{leftSupply} and \code{rightSupply} split this set
into two disjoint subsets without the identifier
obtained by \code{thisID}.
There are different implementations available \cite{AugustssonRittriSynek94}
(see below for a simple implementation) and our system
is parametric over concrete implementations of \code{IDSupply}.

When translating Curry to Haskell, KiCS2 adds to each operation
an additional argument of type \code{IDSupply}.
For instance, the operation \code{aBool}
defined in Sect.~\ref{ex:aBool} is translated into:
\begin{haskell}
  aBool :: IDSupply -> Bool
  aBool s = Choice (thisID s) True False
\end{haskell}
Similarly, the operation
\begin{curry}
  main :: Bool
  main = xorSelf aBool
\end{curry}
is translated into
\begin{haskell}
  main :: IDSupply -> Bool
  main s = xorSelf (aBool (leftSupply s)) (rightSupply s)
\end{haskell}
so that the set \code{s} is split into a set \code{(leftSupply s)}
containing identifiers for the evaluation of \code{aBool}
and a set \code{(rightSupply s)} containing identifiers
for the evaluation of the operation \code{xorSelf}.

Since all data types are extended by additional constructors,
we must also extend the definition of operations performing
pattern matching.\footnote{To obtain a simple compilation scheme,
KiCS2 transforms source programs into uniform programs
\cite{BrasselHanusPeemoellerReck11} where pattern matching
is restricted to a single argument.
This is always possible by introducing auxiliary operations.}
For instance, consider the definition of polymorphic lists
\begin{curry}
  data List a = Nil | Cons a (List a)
\end{curry}
and an operation to extract the first element of a non-empty list:
\begin{curry}
  head :: List a -> a
  head (Cons x xs) = x
\end{curry}
The type definition is then extended as follows:
\begin{haskell}
  data List a = Nil | Cons a (List a) | Choice ID (List a) (List a) | Fail
\end{haskell}
The operation \code{head} is extended by an identifier supply
and further matching rules:
\begin{haskell}
  head :: List a -> IDSupply -> a
  head (Cons x xs)      s = x
  head (Choice i x1 x2) s = Choice i (head x1 s) (head x2 s)
  head _                s = Fail
\end{haskell}
The second rule transforms a non-deterministic argument
into a non-deterministic result and
the final rule returns \code{Fail} in all other cases,
i.e., if \code{head} is
applied to the empty list as well as if the matching argument
is already a failed computation (failure propagation).

To show a concrete example, we use the following implementation
of \code{IDSupply} based on unbounded integers:
\begin{haskell}
  type IDSupply = Integer
  initSupply    = return 1
  thisID      n = n
  leftSupply  n = 2 * n
  rightSupply n = 2 * n + 1
\end{haskell}
If we apply the same transformation to the rules defining \code{xor}
and evaluate the main expression \code{(main 1)},
we obtain the result
\begin{haskell}
  Choice 2 (Choice 2 False True) (Choice 2 True False)
\end{haskell}
Thus, the result is non-deterministic and contains three choices,
whereby all of them have the same identifier.
To extract all values from such a \code{Choice} structure,
we have to traverse it and  compute all possible choices
but consider the choice identifiers to make consistent (left/right)
decisions.
Thus, if we select the left branch as the value
of the outermost \code{Choice}, we also have to select the left branch
in the selected argument \code{(Choice 2 False True)} so that \code{False}
is the only value possible for this branch.
Similarly, if we select the right branch as the value of the outermost
\code{Choice}, we also have to select the right branch in
its selected argument \code{(Choice 2 True False)}, which again yields 
\code{False} as the only possible value. In consequence, 
the unintended value \code{True} is not produced.

The requirement to make consistent decisions can be implemented by storing
the decisions already made for some choices during the traversal.
For this purpose, we introduce the type
\begin{haskell}
  data Decision = NoDecision | ChooseLeft | ChooseRight
\end{haskell}
where \code{NoDecision} represents the fact that the value of a choice has
not been decided yet. Furthermore, we assume operations to lookup the 
current decision for a given identifier or change it
(depending on the implementation
of \code{IDSupply}, KiCS2 supports several implementations
based on memory cells or finite maps):
\begin{haskell}
  lookupDecision :: ID -> IO Decision
  setDecision    :: ID -> Decision -> IO ()
\end{haskell}
Now we can print all values contained in a choice structure
in a depth-first manner by the following I/O operation:\footnote{%
Note that this code has been simplified for readability
since the type system of Haskell does not
allow this direct definition.}
\label{sec:printValsDFS}
\begin{haskell}
  printValsDFS :: a -> IO ()$\listline$
  printValsDFS Fail             = return ()$\listline$
  printValsDFS (Choice i x1 x2) = lookupDecision i >>= follow
    where
      follow ChooseLeft  = printValsDFS x1
      follow ChooseRight = printValsDFS x2
      follow NoDecision  = do newDecision ChooseLeft  x1
                              newDecision ChooseRight x2$\listline$
      newDecision d x = do setDecision i d
                           printValsDFS x
                           setDecision i NoDecision$\listline$
  printValsDFS v = print v
\end{haskell}
This operation ignores failures and prints values that are not rooted by a
\code{Choice} constructor. For a \code{Choice} constructor, 
it checks whether a decision for
this identifier has already been made (note that the initial value
for all identifiers is \code{NoDecision}).
If a decision has been made for this choice, it follows this decision.
Otherwise, the left alternative is used and this decision is stored.
After printing all values w.r.t.\ this decision,
the decision is undone (like in backtracking)
and the right alternative is used and stored.

In general, this operation is applied to the normal form
of the main expression (where \code{initSupply} is used to
compute an initial identifier supply passed to this expression).
The normal form computation is necessary for structured data like lists,
so that a failure or choice in some part of the data is moved to
the root.

Other search strategies, like
breadth-first search, iterative deepening, or parallel search,
can be obtained by different implementations of this main operation
to print all values.
Furthermore, one can also collect all values in a tree-like data structure
so that the programmer can implement his own search strategies
(this corresponds to encapsulating search \cite{BrasselHanusHuch04JFLP}).
Finally, instead of printing all values, one can easily define operations
to print either the first solution only or one by one upon user request.
Due to the lazy evaluation strategy of Haskell,
such operations can also be applied to infinite choice structures.

To avoid an unnecessary growth of the search space represented by
\code{Choice} constructors, our compiler performs an optimization for
deterministic operations. If an operation is defined by non-overlapping
rules and does not call, neither directly nor indirectly through
other operations, a function defined by overlapping rules,
the evaluation of such an operation (like \code{xor} or \code{not})
cannot introduce non-deterministic values.
Thus, it is not necessary to pass an identifier supply to the operation.
In consequence, only the matching rules are extended by additional cases
for handling \code{Choice} and \code{Fail} so that the generated code
is nearly identical to a corresponding functional program.
Actually, the benchmarks presented in \cite{BrasselHanusPeemoellerReck11}
show that for deterministic operations this implementation outperforms 
all other Curry implementations,
and, for non-deterministic operations, outperforms Prolog-based
implementations of Curry and can compete with MCC \cite{Lux99FLOPS},
a Curry implementation that compiles to C.

As mentioned in the introduction, occurrences of logic variables are
translated into generators.
For instance, the expression \ccode{not x}, where \code{x} is a logic variable,
is translated into \ccode{not (aBool s)}, where \code{s} is an \code{IDSupply}
provided by the context of the expression. The latter expression is 
evaluated by reducing the
argument \code{aBool s} to a choice between \code{True} or \code{False}
followed by applying \code{not} to this choice.
This is similar to a narrowing step on \ccode{not x}
that instantiates the variable \code{x} to \code{True} or \code{False}.
Since such generators are standard
non-deterministic operations, they are translated like any other operation
and, therefore, do not require any additional run-time support.
However, in the presence of equational constraints,
there are methods which are more efficient than generating all values.
These methods and their implementation are discussed in the
next section.

\section{Equational Constraints and Unification}
\label{sec:Unification}

As known from logic programming, predicates or constraints
are important to restrict the set of intended values
in a non-deterministic computation.
Apart from user-defined predicates,
equational constraints of the form \code{$e_1$\,=:=\,$e_2$}
are the most important kind of constraints.
We have already seen a typical application
of an equational constraint in the operation
\code{last} in Sect.~\ref{ex:last}.

Due to the presence of non-terminating operations
and infinite data structures,
\ccode{=:=} is interpreted as the \emph{strict equality} on terms
\cite{GiovannettiLeviMoisoPalamidessi91},
i.e., the equation \code{$e_1$\,=:=\,$e_2$} is satisfied iff
$e_1$ and $e_2$ are reducible to unifiable constructor terms.
In particular, expressions that do not have a value
are not equal w.r.t.\ \ccode{=:=}, e.g.,
the equational constraint \ccode{head [] =:= head []}
is not satisfiable.\footnote{From now on, we use the
standard notation for lists, i.e., \code{[]} denotes the empty list
and \code{(x:xs)} denotes a list with head element \code{x}
and tail \code{xs}.}

Due to this constructive definition, \ccode{=:=}
can be considered as a binary function defined by the
following rules (we only present the rules for the Boolean
and list types, where \code{Success} denotes the only constructor
of the type \code{Success} of constraints):
\begin{curry}
  True   =:= True    =  Success
  False  =:= False   =  Success

  []     =:= []      =  Success
  (x:xs) =:= (y:ys)  =  x =:= y & xs =:= ys

  Success & c  =  c
\end{curry}
If we translate these operations into Haskell by the scheme
presented in Sect.~\ref{sec:Compilation},
the following rules are added to these rules
in order to propagate choices and failures:
\begin{haskell}
  Fail         =:= _             =  Fail
  _            =:= Fail          =  Fail
  Choice i l r =:= y             =  Choice i (l =:= y) (r =:= y)
  x            =:= Choice i l r  =  Choice i (x =:= l) (x =:= r)
  _            =:= _             =  Fail

  Fail         & _     =  Fail
  Choice i l r & c     =  Choice i (l & c) (r & c)
  _            & _     =  Fail

\end{haskell}
Although this is a correct implementation of equational constraints,
it might lead to an unnecessarily large search space
when it is applied to generators representing logic variables.
For instance, consider the following generator for Boolean lists:
\begin{curry}
  aBoolList = [] ? (aBool : aBoolList)
\end{curry}
This is translated into Haskell as follows:
\begin{haskell}
  aBoolList :: IDSupply -> [Bool]
  aBoolList s = Choice (thisID s) [] (aBool (leftSupply s)
                                      : aBoolList (rightSupply s))
\end{haskell}
Now consider the equational constraint \ccode{x =:= [True]}.
If the logic variable \code{x} is replaced by \code{aBoolList},
the translated expression \ccode{aBoolList s =:= [True]}
creates a search space when evaluating its first argument,
although there is no search required since there is only one
binding for \code{x} satisfying the constraint.
Furthermore and even worse, unifying two logic variables
introduces an infinite search space. For instance,
the expression \ccode{xs =:= ys \& xs++ys =:= [True]}
results in an infinite search space when the logic variables
\code{xs} and \code{ys} are replaced by generators.

To avoid these problems, we have to implement the idea
of the well-known unification principle \cite{Robinson65}.
Instead of enumerating all values for logic variables
occurring in an equational constraint,
we \emph{bind} the variables to another variable or term.
Since we compile into a purely functional language,
the binding cannot be performed by some side effect.
Instead, we add binding constraints to the computed results to be 
processed by a search strategy that extracts values from choice structures.

To implement unification, we have to distinguish free variables
from ``standard choices'' (introduced by overlapping rules)
in the target code. For this purpose, we refine the definition
of the type \code{ID} as follows:\footnote{For the sake
of simplicity, in the following, we consider the
implementation of \code{IDSupply} to be unbounded integers.}
\begin{haskell}
  data ID = ChoiceID Integer | FreeID Integer
\end{haskell}
The new constructor \code{FreeID} identifies a choice corresponding to
a free variable, e.g., the generator for Boolean variables is
redefined as
\begin{haskell}
  aBool s = Choice (FreeID (thisID s)) True False
\end{haskell}
If an operation is applied to a free variable and requires its value,
the free variable is transformed into a standard choice.
For this purpose, we define a simple operation to perform
this transformation:
\begin{haskell}
  narrow :: ID -> ID
  narrow (FreeID i) = ChoiceID i
  narrow x          = x
\end{haskell}
Furthermore, we use this operation in narrowing steps,
i.e., in all rules operating on \code{Choice} constructors.
For instance, in the implementation of the operation \code{not}
we replace the rule
\begin{haskell}
  not (Choice i x1 x2) s = Choice i (not x1 s) (not x2 s)
\end{haskell}
by the rule
\begin{haskell}
  not (Choice i x1 x2) s = Choice (narrow i) (not x1 s) (not x2 s)
\end{haskell}
As mentioned above, the consideration of free variables
is relevant in equational constraints where \emph{binding constraints} are
generated. For this purpose, we introduce a type to represent
a binding constraint as a pair of a choice identifier
and a decision for this identifier:
\begin{haskell}
  data Constraint = ID :=: Decision
\end{haskell}
Furthermore, we extend each data type by the possibility to
add constraints:
\begin{haskell}
  data Bool   = $\ldots$ | Guard [Constraint] Bool
  data List a = $\ldots$ | Guard [Constraint] (List a)
\end{haskell}
A single \code{Constraint} provides the decision for
one constructor. In order to support constraints for structured data,
a list of \code{Constraint}s provides the decision for the outermost
constructor and the decisions for all its arguments.
Thus, \code{(Guard $cs$ $v$)} represents a \emph{constrained value},
i.e., the value $v$ is only valid if the constraints $cs$ are
consistent with the decisions previously made during search.
These binding constraints are created by the
equational constraint operation \ccode{=:=}: if a free variable
should be bound to a constructor, we make the same
decisions as it would be done in the successful branch
of the generator. In case of Boolean values,
this can be implemented by the following additional rules
for \ccode{=:=}:
\begin{haskell}
  Choice (FreeID i) _ _ =:= True   =  Guard [i :=: ChooseLeft ] Success
  Choice (FreeID i) _ _ =:= False  =  Guard [i :=: ChooseRight] Success
\end{haskell}
Hence, the binding of a variable to some known value
is implemented as a binding constraint for the choice identifier
for this variable. However, if we want to bind a variable
to another variable, we cannot store a concrete decision.
Instead, we store the information that the decisions for
both variables, when they are made to extract values,
must be identical. For this purpose, we extend the \code{Decision}
type to cover this information:
\begin{haskell}
  data Decision = $\ldots$ | BindTo ID
\end{haskell}
Furthermore, we add the rule that an equational constraint
between two variables yields a binding for these variables:%
\begin{haskell}
  Choice (FreeID i) _ _ =:= Choice (FreeID j) _ _
    =  Guard [i :=: BindTo j] Success
\end{haskell}
The consistency of constraints is checked when values are extracted
from a choice structure, e.g., by the operation \code{printValsDFS}.
For this purpose, we extend the definition of the corresponding search
operations by calling a constraint solver for the constraints.
For instance, the definition of \code{printValsDFS} is extended by
a rule handling constrained values:
\begin{haskell}
  $\ldots$
  printValsDFS (Guard cs x) = do consistent <- add cs
                                 if consistent then do printValsDFS x
                                                       remove cs
                                               else return ()
  $\ldots$
\end{haskell}
The operation \code{add} checks the consistency of the constraints \code{cs}
with the decisions made so far and, in case of consistency,
stores the decisions made by the constraints.
In this case, the constrained value is evaluated
before the constraints are removed to allow backtracking.
Furthermore, the operations \code{lookupDecision} and \code{setDecision}
are extended to deal with bindings between two variables,
i.e., they follow variable chains in case of \code{BindTo} constructors.

Finally, with the ability to distinguish free variables 
(choices with an identifier of the form \code{(FreeID $\ldots$)})
from other values during search, values containing logic variables 
can also be printed in a specific form rather than enumerating all values, 
similarly to logic programming systems. For instance, KiCS2 evaluates the 
application of \code{head} to an unknown list as follows:
\begin{haskell}
  Prelude> head xs where xs free
  {xs = (_x2:_x3)} _x2
\end{haskell}
Here, free variables are marked by the prefix \code{\us{}x}.

\section{Functional Patterns}
\label{sec:FuncPatterns}

A well-known disadvantage of equational constraints
is the fact that \ccode{=:=} is interpreted as strict equality.
Thus, if one uses equational constraints to express
requirements on arguments, the resulting operations might be too strict.
For instance, the equational constraint in the condition
defining \code{last} (see Sect.~\ref{ex:last})
requires that \code{ys++[e]} as well as \code{xs}
must be reducible to unifiable terms so that in consequence
the input list \code{xs} is completely evaluated.
Hence, if \code{failed} denotes an operation whose evaluation fails,
the evaluation of \code{last [failed,True]} has no result.
On the other hand, the evaluation of \code{last' [failed,True]} yields
the value \code{True}, i.e., the definition of \code{last'} is less strict
thanks to the use of functional patterns.

As another example for the advantage of the reduced strictness
implied by functional patterns,
consider an operation that returns the first duplicate element in
a list.
Using equational constraints, we can define it as follows:
\begin{curry}
  fstDup xs | xs =:= ys++[e]++zs & elem e ys =:= True & nub ys =:= ys
            = e    where ys, zs, e free
\end{curry}
The first equational constraint is used to split the input list \code{xs}
into three sublists.
The last equational constraint ensures that the first sublist
\code{ys} does not contain duplicated elements
(the library operation \code{nub} removes all duplicates from a list)
and the second equational constraint ensures that the first element
after \code{ys} occurs in \code{ys}.
Although this implementation is concise, it cannot be applied
to infinite lists due to the strict interpretation of \ccode{=:=}.
This is not the case if we define this operation by a functional pattern:
\begin{curry}
  fstDup' (ys++[e]++zs) | elem e ys =:= True & nub ys =:= ys
                        = e
\end{curry}
Because of the reduced strictness, the logic variable \code{zs} (matching
the tail list after the first duplicate) is never evaluated.
This is due to the fact that a functional pattern like
\code{(xs++[e])} abbreviates all values to which it can be evaluated
(by narrowing), like \code{[e]}, \code{[x1,e]}, \code{[x1,x2,e]} etc.
Conceptually, the rule defining \code{last'}
abbreviates the following (infinite) set of rules:
\begin{curry}
  last' [e] = e
  last' [x1,e] = e
  last' [x1,x2,e] = e
  $\ldots$
\end{curry}
Obviously, one cannot implement functional patterns by
a transformation into an infinite set of rules. Instead, they are
implemented by a specific \emph{lazy unification} procedure \ccode{=:<=}
\cite{AntoyHanus05LOPSTR}.
For instance, the definition of \code{last'} is transformed into
\begin{curry}
  last' ys | (xs++[e]) =:<= ys  = e   where xs, e free
\end{curry}
The behavior of \ccode{=:<=} is similar to \ccode{=:=},
except for the case that a variable in the left argument
should be bound to some expression: instead of evaluating
the expression to some value and binding the variable to the value,
the variable is bound to the \emph{unevaluated} expression
(see \cite{AntoyHanus05LOPSTR} for more details).
Due to this slight change, failures or infinite structures
in actual arguments do not cause problems in the matching
of functional patterns.

The general structure of the implementation of functional patterns in KiCS2
is quite similar to that of equational constraints,
with the exception that variables could be also bound
to unevaluated expressions. 
Only if such variables are later
accessed, the expressions they are bound to are evaluated.
This can be achieved by adding a further alternative
to the type of decisions:
\begin{haskell}
  data Decision = $\ldots$ | LazyBind [Constraint]
\end{haskell}
The implementation of the lazy unification operation \ccode{=:<=}
is almost identical to the strict unification operation \ccode{=:=}
as shown in Sect.~\ref{sec:Unification}.
The only difference is in the rules where a free variable occurs
in the left argument. All these rules are replaced by the single
rule
\begin{haskell}
  Choice (FreeID i) _ _ =:<= x
    = Guard [i :=: LazyBind (lazyBind i x)] Success
\end{haskell}
where the auxiliary operation \code{lazyBind}
implements the demand-driven evaluation of the right argument \code{x}:
\begin{haskell}
  lazyBind :: ID -> a -> [Constraint]
  lazyBind i True  = [i :=: ChooseLeft]
  lazyBind i False = [i :=: ChooseRight]
\end{haskell}
The use of the additional \code{LazyBind} constructor allows the argument 
\code{x} to be stored in a binding constraint without evaluation 
(due to the lazy evaluation strategy of the
target language Haskell).
However, it is evaluated by \code{lazyBind} when its binding
is required by another part of the computation.
Similarly to equational constraints,
lazy bindings are processed by a solver when values are extracted.
In particular, if a variable has more than one lazy binding constraint
(which is possible if a functional pattern evaluates to a non-linear term),
the corresponding expressions are evaluated and unified
according to the semantics of functional patterns
\cite{AntoyHanus05LOPSTR}.

In order to demonstrate the operational behavior of our implementation,
we sketch the evaluation of the lazy unification constraint
\code{xs++[e] =:<= [failed,True]} that occurs when the expression
\code{last' [failed,True]} is evaluated (we omit failed branches
and some other details; note that logic variables are replaced
by generators, i.e., we assume that \code{xs} is replaced by
\code{aBoolList 2} and \code{e} is replaced by \code{aBool 3}):
\begin{haskell}
     aBoolList 2 ++ [aBool 3] =:<= [failed, True]
  $\leadsto$ [aBool 4, aBool 3] =:<= [failed, True]
  $\leadsto$ aBool 4 =:<= failed & aBool 3 =:<= True & [] =:<= []
  $\leadsto$ Guard [ 4 :=: LazyBind (lazyBind 4 failed)
           , 3 :=: LazyBind (lazyBind 3 True)] Success
\end{haskell}
If the value of the expression \code{last' [failed,True]} is later required,
the value of the variable \code{e} (with the identifier \code{3}) is in turn
required.
Thus, \code{(lazyBind 3 True)} is evaluated to \code{[3 :=: ChooseLeft]}
which corresponds to the value \code{True} of the generator \code{(aBool 3)}.
Note that the variable with identifier \code{4} does not occur
anywhere else, so that the binding \code{(lazyBind 4 failed)}
will never be evaluated, as intended.

\section{Benchmarks}
\label{sec:Benchmarks}

In this section we evaluate our implementation of
equational constraints and functional patterns by some benchmarks.
The benchmarks were executed on a Linux machine running Debian 5.0.7 with
an Intel Core 2 Duo (3.0GHz) processor.
KiCS2 has been used with
the Glasgow Haskell Compiler (GHC 7.0.4, option -O2) as its backend and
an efficient \code{IDSupply} implementation that makes use of \code{IORef}s.
For a comparison with other mature implementations of Curry,
we considered PAKCS \cite{Hanus10PAKCS}
(version 1.9.2, based on a SICStus-Prolog 4.1.2)
and MCC \cite{Lux99FLOPS} (version 0.9.10). The timings were performed with the
time command measuring the execution time (in seconds) of a compiled
executable for each benchmark as a mean of three runs.
The programs used for the benchmarks,
partially taken from \cite{AntoyHanus05LOPSTR},
are \code{last} (compute the last element of a list),\footnote{%
\ccode{inc x n} is a naive addition that \code{n} times
increases its argument \code{x} by 1.}
\code{simplify} (simplify a symbolic arithmetic expression),
\code{varInExp} (non-deterministically return a variable occuring
                  in a symbolic arithmetic expression),
\code{half} (compute the half of a Peano number using logic variables),
\code{palindrome} (check whether a list is a palindrome),
\code{horseman} (solving an equation relating heads and feet of horses and men
based on Peano numbers),
and
\code{grep} (string matching based on a non-deterministic
specification of regular expressions \cite{AntoyHanus10CACM}).

\begin{figure}[t]
\centering
\begin{tabular}{|l|r|r|r|}
\hline
Expression                              & \code{==~}& \code{=:=}& \code{=:<=} \\
\hline
\code{last (map (inc 0) [1..10000])}   &  2.91   &    0.05   &    0.01   \\
\code{simplify}                         & 10.30      & 6.77      & 7.07        \\
\code{varInExp}                         & 2.34    & 0.24      & 0.21        \\
\code{fromPeano (half (toPeano 10000))} & 26.67     & 5.95      & 11.19       \\
\code{palindrome}                       & 30.86     & 14.05     & 20.26       \\
\code{horseman}                         & 3.24      & 3.31      & n/a         \\
\code{grep}                             & 1.06      & 0.10      & n/a         \\
\hline
\end{tabular}
\caption{Benchmarks: comparing different representations for equations}
 \label{fig:search-strategies}
\end{figure}

In Sect.~\ref{sec:Unification} we mentioned that equational
constraints could also be solved by generators without variable bindings,
but this technique might increase the search space due to the
possibly superfluous generation of all values.
To show the beneficial effects of our implementation
of equational constraints with variable bindings,
in Fig.~\ref{fig:search-strategies} we compare
the results of using equational constraints (\code{=:=})
to the results
where the Boolean equality operator (\code{==}) is used (which
does not perform bindings but enumerate all values).
As expected, in most cases the creation and traversal of a large
search space introduced by \code{==} is much slower than our
presented approach with variable bindings.
In addition, the example \code{last} shows that the lazy 
unification operator (\ccode{=:<=}) improves the performance when unifying an
expression which has to be evaluted only partially. Using strict unification,
all elements of the list are (unnecessarily) evaluated.

\begin{figure}[t]
\centering
\begin{tabular}{|l|r|r|r|}
\hline
Expression                              & KiCS2 & PAKCS  & MCC  \\
\hline
\code{last (map (inc 0) [1..10000])}   & 0.05  & 0.40   &  0.01\\
\code{simplify}                         & 6.77  & 0.15   & 0.00 \\
\code{varInExp}                         & 0.24  & 0.89   & 0.07 \\
\code{fromPeano (half (toPeano 10000))} & 5.95  & 108.88 & 3.22 \\
\code{palindrome}                       & 14.05 & 32.56  & 1.07 \\
\code{horseman}                         & 3.31  & 8.70   & 0.42 \\
\code{grep}                             & 0.10  & 2.88   & 0.14 \\
\hline
\end{tabular}
\caption{Benchmarks: strict unification in different Curry implementations}
 \label{fig:unification}
\end{figure}

In contrast to the Curry implementations PAKCS and MCC,
our implementation of strict unification is based
on an explicit representation of the search space instead of
backtracking and manipulating
a global state containing bindings for logic variables.
Nevertheless, the benchmarks in Fig.~\ref{fig:unification},
using equational constraints only,
show that it can compete with or even outperform the other implementations.
The results show that the implementation of unification
of MCC performs best. However, in most cases our implementation 
outperforms the Prolog-based PAKCS implementation, 
except for some examples. In particular, \code{simplify} does not perform well
due to expensive bindings of free variables to large arithmetic expressions
in unsuccessful branches of the search. Further investigation
and optimization will hopefully lead to a better performance in
such cases.

\begin{figure}[t]
\centering
\begin{tabular}{|l|r|r|}
\hline
Expression                              & KiCS2 & PAKCS    \\
\hline
\code{last (map (inc 0) [1..10000])}   &  0.01 & 0.33     \\
\code{simplify}                         & 7.07  & 0.27     \\
\code{varInExp}                         &  0.21  &  1.87    \\
\code{fromPeano (half (toPeano 10000))} & 11.19 & $\infty$ \\
\code{palindrome}                       & 20.26 & $\infty$ \\
\hline
\end{tabular}
\caption{Benchmarks: functional patterns in different Curry implementations}
 \label{fig:functional-patterns}
\end{figure}

As MCC does not support functional patterns, the performance of lazy
unification is compared with PAKCS only (Fig.~\ref{fig:functional-patterns}).
Again, our compiler performs well against PAKCS
and outperforms it in most cases (``$\infty$'' denotes a run time
of more than 30 minutes).

\section{Conclusions and Related Work}
\label{sec:Conclusions}

We have presented an implementation of equational constraints
and functional patterns in KiCS2, a purely functional implementation
of Curry. Our implementation is based on adding binding constraints
to computed values and processing them when values are extracted
at the top level of a computation.
Since we only have added new constructors and pattern matching rules 
for them in our implementation, no overhead is introduced for programs
without equational constraints, i.e.,
our implementation does not sacrifice the high efficiency
of the kernel implementation shown in
\cite{BrasselHanusPeemoellerReck11}.
However, if these features are used, they usually lead
to a comparably efficient execution, as demonstrated by our benchmarks.

Other implementations of equational constraints in functional logic
languages use side effects for their implementation.
For instance, PAKCS \cite{Hanus10PAKCS}
exploits the implementation of logic variables in Prolog,
which are implemented on the primitive level by side effects.
MCC \cite{Lux99FLOPS} compiles into C where a specific
abstract machine implements the handling of logic variables.
We have shown that our implementation is competitive to those.
In contrast to those systems, our implementation supports
a variety of search strategies, like breadth-first or parallel search,
where the avoidance of side effects is important.

For future work it might be interesting to add
further constraint structures to our implementation, like real arithmetic
or finite domain constraints.
This might be possible by extending the kinds of constraints
of our implementation and solving them by
functional approaches like \cite{SchrijversStuckeyWadler09}.

\end{document}